\documentclass[ASNA,twocolumn]{USG} 
\usepackage{anyfontsize} %

\usepackage{braket}

\usepackage{steinmetz}
\graphicspath{{./images/}}

\articletype{RESEARCH ARTICLE}%

\received{XX Month Year}
\revised{XX Month Year}
\accepted{XX Month Year}
\journal{Journal Name}
\volume{0}
\copyyear{Year}
\startpage{1}
\articledoi{10.1002/0000}

\begin{document}
\title{Phase Sensitivity of Spectrally Multimode SU(1,1) Interferometers with Waveguide based Optical Parametric Amplifier}
\transtitle{Phase Sensitivity of Multimode SU(1,1) Interferometers with Waveguide based Optical Parametric Amplifier}
\subtranstitle{trans-subtitle}
\author[1]{Sonu Jana}[https://orcid.org/0009-0005-1798-2594]
\author[1]{Swagatam Bag}[https://orcid.org/0009-0005-0325-6970]
\author[1,2]{Fabien Bretenaker}
\author[3]{Nadia Belabas}
\author[1]{Syamsundar De}

\authormark{TAYLOR \textsc{et al.}}
\titlemark{PLEASE INSERT YOUR ARTICLE TITLE HERE}

\address[1]{\orgdiv{Centre for Interdisciplinary and Convergent Technologies, }\orgname{Indian Institute of Technology Kharagpur,}%
\orgaddress{\state{West Bengal, }\country{India}}}

\address[2]{\orgdiv{LuMin, }\orgname{Universit\'e Paris-Saclay, CNRS, ENS Paris-Saclay, CentraleSup\'elec, }%
\orgaddress{\state{Orsay, }\country{France}}}

\address[3]{\orgdiv{Centre for Nanosciences and Nanotechnologies, CNRS,}\orgname{Universit\'e Paris-Saclay,}%
\orgaddress{\state{Palaiseau, }\country{France}}}


\corres{Syamsundar De  (\email{syamsundarde@atdc.iitkgp.ac.in}) ~|~ Sonu Jana (\email{sonujana@kgpian.iitkgp.ac.in})}


\presentaddress{This is sample for present address text this is sample for present address text.}



\transkeywords{Keyword1 | Keyword2 | Keyword3 | Keyword4 | Keyword5}

\abstract[ABSTRACT]{We present a comprehensive framework for evaluating the phase sensitivity of spectrally multimode SU(1,1) interferometers probed with a coherent-vacuum input, $\ket{\alpha_{f}}_{s} \otimes \ket{0}_{i} $, under both number and homodyne detections. The optical parametric amplifiers (OPAs) are simulated using a realistic periodically polled thin-film lithium niobate (PP-TFLN) waveguide. The theoretical model incorporates the intrinsic multimode spectral nature of waveguide-based OPAs. Under the assumptions of identical OPAs and Schmidt-mode-independent phase shifts, we have shown that the multimode SU(1,1) interferometer is equivalent to a collection of independent single-mode interferometers in the Schmidt basis that operate in parallel, each associated with a Schmidt mode of the parametric process. We further present an optimized waveguide design based on asymmetric group-velocity matching to realize a practical OPA with nearly factorizable joint spectral amplitude (JSA). The Schmidt coefficients extracted from the numerically simulated JSA are incorporated into the theoretical model to evaluate the phase sensitivity. The results show that the sensitivity under a multimode condition depends strongly on the measurement scheme. For number detection, the degradation in sensitivity arises from the redistribution of the available nonlinear resource among the Schmidt modes, whereas homodyne detection exhibits an additional coherence penalty that can be substantially reduced by optimally shaping the local oscillator. Moreover, we show that the performance degradation can be to a large extent mitigated by optimizing the spectrum of the injected signal coherent state $\ket{\alpha_{f}}_{s}$  and, in the case of homodyne detection, the spectrum of the local oscillator. Altogether, this work provides a unified theoretical and numerical framework for analyzing and optimizing realistic multimode SU(1,1) interferometers with waveguide-based OPAs for quantum-enhanced sensing applications.}



\maketitle


\section{Introduction}\label{sec1}

Optical interferometry is a powerful technique for precision measurements, with applications in biological sensing \cite{1_Gallego2009,1_Jaros2022,1_Jha2024}, inertial navigation \cite{2_DellOlio2023,2_Song2023}, spectroscopy \cite{3_Chen2022,3_Yang2021}, and fundamental physics \cite{4_Abbott2016,4_Akutsu2019}. Most conventional interferometric schemes employ classical states of light \cite{5_Groot2019,5_Yang2018}, for which the achievable phase sensitivity is limited by the shot noise \cite{6_Caves1981,6_Takeoka2017}. To surpass this classical bound, numerous non-classical states of light \cite{7_Lawrie2019,7_Tan2014,7_Gietka2017} and novel interferometric setups \cite{8_Du2022,8_Li2023,8_Liu2023,8_Natan2026,8_Szigeti2017} have been proposed and investigated. Among the various interferometric architectures, the SU(1,1) / nonlinear interferometer \cite{9_Yurke1986} has emerged as a promising platform for achieving phase sensitivity beyond the shot-noise limit even with a classical state of light \cite{10_Plick2010}. In contrast to the conventional Mach-Zehnder interferometer (MZI), the SU(1,1) interferometer replaces passive beam splitters with OPAs, which serve as beam-splitting and recombining elements. The first OPA generates the quantum correlations required to achieve sensitivity beyond the shot noise limit \cite{10_Plick2010,11_Ou2020,11_Salykina2023}. Due to the possibility of achieving sub-shot-noise sensitivity with classical light inputs, the SU(1,1) interferometer has been the subject of extensive theoretical investigation \cite{10_Plick2010,12_Giese2017,12_Hu2016,12_Li2014,12_Xu2022,12_Zhang2023}. At the same time, several proof-of-principle experiments have demonstrated its potential for quantum-enhanced phase estimation \cite{13_Hudelist2014,13_Frascella2019,13_Li2021,13_Liu2019,13_Lukens2018}. In addition, SU(1,1) interferometers exhibit an inherent robustness to detection losses compared to their MZI counterparts \cite{14_Ou2012,14_Marino2012,14_Manceau2017,14_Jana2026}, making them attractive candidates for practical quantum sensing.

Most SU(1,1) interferometers have been experimentally realized using various nonlinear platforms, including parametric down-conversion (PDC) in $\chi^{(2)}$ nonlinear crystals \cite{13_Frascella2019} and four-wave mixing (FWM) based on  $\chi^{(3)}$ nonlinearity in atomic vapors \cite{Du2018,13_Hudelist2014} and optical fibers \cite{13_Li2021,13_Liu2019,13_Lukens2018}. However, recent advances in integrated nonlinear photonics have enabled the realization of the OPA for the SU(1,1) interferometer with compact waveguide technologies \cite{15_Chen2025,15_Ledezma2022,15_Peng2025}. 
Furthermore, recent theoretical studies have explored the realization of a fully integrated SU(1,1) interferometer using periodically poled potassium titanylphosphate (PPKTP) waveguides \cite{16_Ferreri2021}. These advances established the waveguide-based OPA as a promising route toward the practical realization of SU(1,1) interferometry. Among the various $\chi^{(2)}$-based integrated photonic platforms, periodically poled thin-film lithium niobate (PP-TFLN) waveguides have emerged as a promising platform due to their large second-order nonlinearity ($d_{33} = 25\,\mathrm{pm/V}$), ultra-low propagation losses  (down to $0.02\,\mathrm{dB/cm}$ in state-of the fabricated waveguides), strong optical confinement ($\mathrm{refractive\, index} \approx 2.2 $), a broad optical transparency window ($350\,\mathrm{nm}-5000\,\mathrm{nm}$), and a mature fabrication technology \cite{17_Lin2020,17_Qi2020,17_Stokowski2023,17_Zhu2021}. These attributes enable highly efficient nonlinear interactions over compact device lengths, making PP-TFLN an attractive platform for waveguide-based OPAs for SU(1,1) interferometers. In a PP-TFLN waveguide-based OPA, the generated quantum state is governed by the PDC process, which exhibits spectrally multimode behavior under Schmidt decomposition due to finite pump bandwidth, waveguide dispersion, and quasi-phase-matching (QPM) conditions, especially in the ultrafast-pulsed regime \cite{18_Laudenbach2016,18_RomanRodriguez2021,18_Shi2024}. This intrinsically multimode nature of waveguide-based SU(1,1) interferometers raises several new questions compared with the single-mode case: how is the available gain distributed among the system's Schmidt modes? How does this affect the ultimate phase sensitivity of the interferometer? What is the optimal detection scheme, direct (number) detection or balanced homodyne detection? In this latter case, to which Schmidt mode(s) should the local oscillator be spectrally shaped? In the case where the interferometer is probed with a coherent state, to which mode should this coherent state be adapted? What is the influence of the vacuum that will be injected into the other Schmidt modes? 

To answer these questions, one needs to take into account the nontrivial joint spectral amplitude (JSA) of the PDC process, which completely describes its spectral correlations and its multimode structure, and governs the spectral properties of the two-mode parametric gain. The spectral properties of spontaneous parametric down-conversion (SPDC) sources, including their brightness, purity, and Schmidt-mode structure, have been extensively investigated due to their importance in quantum light generation \cite{18_Laudenbach2016,18_RomanRodriguez2021,18_Shi2024,QL_3_Wang2024}. However, considerably less attention has been devoted to understanding how the intrinsic multimode nature of a realistic waveguide-based OPA influences the phase sensitivity of SU(1,1) interferometers when these sources are employed as the nonlinear elements. Most theoretical analyses of SU(1,1) interferometers continue to rely on an idealized single-mode description of the OPAs, overlooking the influence of the multimode structure on the achievable phase sensitivity. One notable exception is Ref.\,\cite{16_Ferreri2021} that considers the multimode nature of PPKTP-based SU(1,1) interferometers, but mostly in the case where the system is seeded with vacuum. Here, we focus on a PP-TFLN waveguide-based interferometer in which the signal is seeded with a coherent state, and the idler mode is in vacuum. In this situation, to the best of our knowledge, no comprehensive analysis of how the intrinsic multimode structure of waveguide-based OPAs affects the phase sensitivity of the SU(1,1) interferometer has yet been reported. As nonlinear interferometers move toward practical implementation \cite{app_Cardoso2024,app_Dong2025}, it becomes essential to understand how the multimode structure of the OPAs affects achievable phase sensitivity and how their spectral properties can be engineered to optimize interferometric performance. 

In this work, we thus develop a comprehensive theoretical framework for sensitivity evaluation for a realistic spectrally multimode SU(1,1) interferometer by incorporating the intrinsic multimode nature of the waveguide-based OPA through Schmidt-mode representation of the PDC process. Under the assumptions of identical OPAs and a mode-independent phase shift, we show that the multimode SU(1,1) interferometers can be rigorously described as a tensor product of independent single-mode interferometers in the Schmidt basis, providing a convenient framework for analyzing realistic interferometers. Within this formalism, we investigate the phase sensitivity in the widely studied case where the interferometer is seeded with a coherent state for the signal and a vacuum state for the idler. Furthermore, we consider both photon-number detection and homodyne detection, while allowing the coherent probe to occupy an arbitrary normalized spectral mode. Our aim is to derive an analytical expression for the sensitivity of the above input condition under the said detection schemes and explore how the intrinsic multimodal structure degrades the achievable sensitivity relative to the ideal single-mode case. Furthermore, we analyze how the origin of this degradation depends on the detection scheme, and how the sensitivity degradation due to the multimode nature can be partially mitigated by spectrally shaping the input signal coherent state and/or by optimizing the local oscillator spectrum in the homodyne detection case.

To demonstrate the practical relevance of the proposed framework, we then apply it to a realistic SU(1,1) interferometer using a PP-TFLN waveguide as an OPA. We present an optimized waveguide design that enables asymmetric group-velocity matching to engineer the PDC process and generate a nearly factorizable JSA with tunable multimodal characteristics. The resulting Schmidt-mode distribution is then incorporated into our sensitivity analysis to establish a direct connection between realistic waveguide design and interferometric performance. In particular, we examine the sensitivity degradation under two experimentally relevant resource constraints: a fixed total nonlinear gain and a fixed total spontaneous photon number, if the interferometer were fed with vacuum. Our results estimate how the multimode penalty decreases as the generated state approaches the single-mode limit, providing quantitative design guidelines for engineering waveguide-based OPAs and for analyzing the performance of a realistic spectrally multimode SU(1,1) interferometer.

\begin{figure*}[ht]
\centering
\includegraphics[width=1.0\linewidth]{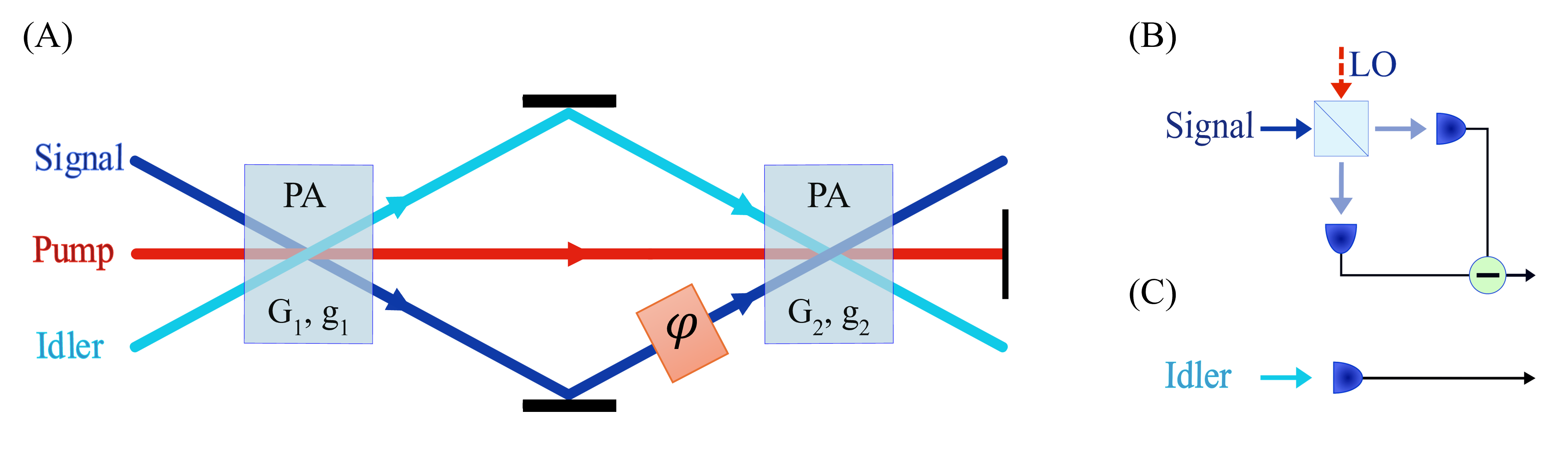}
\caption{(A) Schematic of the general SU(1,1) interferometer considered in this work. The interferometer consists of two optical parametric amplifiers (PAs) with gain coefficients $G_{1}$ and $g_1$, and $G_{2}$ and $g_2$, respectively, with $G_j^2=1+g_j^2$ for $j=\{1,2\}$. The device measures an interferometric phase shift $\varphi$ in the signal arm. Phase estimation is performed using (B) homodyne detection at the signal output and (C) number detection at the idler output.} 
\label{fig1_SU11_block_dia}
\end{figure*}
 
\section{Theoretical Formalism}\label{Theory}
We consider a general SU(1,1) interferometer consisting of two OPAs with amplitude gain parameters $G_{1}$ and $g_1$ and $G_{2}$ and $g_2$, respectively, obeying the relations $G_1^2=1+g_1^2$ and $G_2^2=1+g_2^2$. They are separated by a phase shift section in the signal arm, as illustrated schematically in Figure \ref{fig1_SU11_block_dia}-A. Throughout this work, we restrict our analysis to the balanced configuration, in which the two OPAs have identical gains ($G_{1}=G_{2}=G$). The interferometer is biased at the dark fringe operation point ($\varphi =\pi$) to estimate a small phase perturbation $(\delta \varphi)$. Since the OPAs are realized via the PDC process, a realistic description of the interferometer must account for the OPAs' intrinsically multimode spectral structure. We therefore formulate the interferometer in the Schmidt basis defined by the PDC process. We consider the input state to be a coherent state $\ket{\alpha_{f}}_{s}$ described by a normalized spectral mode $f(\omega)$ for the signal input, and a vacuum state $\ket{0}_{i}$ for the idler input. The $\ket{\alpha_{f}}_{s} \otimes \ket{0}_{i} $ input configuration is the most widely studied probe state for the single-mode SU(1,1) interferometer and one that is readily accessible experimentally \cite{10_Plick2010,13_Lukens2018}. Phase estimation is investigated using two experimentally relevant detection schemes: photon-number detection at the idler output (see Figure \ref{fig1_SU11_block_dia}-C) and homodyne detection at the signal output (see Figure \ref{fig1_SU11_block_dia}-B) \cite{11_Ou2020,14_Jana2026}. These measurement configurations are adopted because they are known to provide optimal phase sensitivity for the coherent-vacuum probe in their respective single-mode interferometers, thereby serving as a natural benchmark for quantifying the influence of the intrinsic multimode structure.

\subsection{Multimode SU(1,1) Interferometer}
\label{Multimode_SU(1,1)_Interferometer}
The two-mode squeezing operator describes the evolution of a single OPA in the continuous mode basis of signal and idler fields \cite{18_RomanRodriguez2021} as -
\begin{equation}
    \hat{S}(r) = \exp\left[r\left(\hat{a}^{\dagger}(\omega_{s})\hat{b}^{\dagger}(\omega_{i})\,-\,\hat{a}(\omega_{s})\hat{b}(\omega_{i})\right)\right],
\end{equation}
where $r$ denotes the squeezing parameter, while $\hat{a}(\omega_{s})$ and $\hat{b}(\omega_{i})$ represent the annihilation operators of the signal and idler modes, respectively. Here, we take $r$ to be real for simplicity, without loss of generality. However, the practical implementation of such an OPA based on PDC is hindered by the finite spectral bandwidth of the parametric process, which gives rise to multiple correlated spectral modes. Consequently, each OPA must be described by a multimode squeezing operator. The JSA completely characterizes this multimode structure of such a parametric process \cite{18_Laudenbach2016,18_RomanRodriguez2021,18_Shi2024}, $F(\omega_{s},\omega_{i})$, which admits the Schmidt decomposition
\begin{equation}
\label{schmit_decomposition}
    F(\omega_{s},\omega_{i})= \sum_{k=1}^{N}\lambda_{k}\,  u_k^{(s)}(\omega_{s})\, v_{k}^{(i)}(\omega_{i})\,,
    \qquad
    \mathrm{with}
    \quad
    \sum_{k=1}^{N} \lambda_{k}^{2} = 1
\end{equation}

Here $\{u_{k}^{(s)}(\omega_{s})\}$, and $\{v_{k}^{(i)}(\omega_{i})\}$ form an orthonormal basis (Schmidt basis) for the signal and idler fields, and $\lambda_{k}$ is the Schmidt coefficient. Thus, the interaction Hamiltonian describing the parametric process can be written as \cite{18_RomanRodriguez2021}- 
\begin{equation}
\hat{H}_{I} = i\hbar\Gamma \iint
F(\omega_s,\omega_i)\, \hat{a}^{\dagger}(\omega_s)\,
\hat{b}^{\dagger}(\omega_i)\, d\omega_s\,d\omega_i
+\mathrm{H.c.}
\label{SPDC_Hamiltonian}
\end{equation}
Where $\Gamma$ is the effective nonlinear coupling constant. Expanding the field operators in the Schmidt basis in terms of $\hat{A}_{k}^{(s)}(\omega_{s})$, $\hat{B}_{k}^{(i)}(\omega_{i})$, which annihilate photon from the mode $u_{k}^{(s)}(\omega_{s})$ and $v_{k}^{(i)}(\omega_{i})$ respectively \cite{20_PhysRevX.5.041017,20_Ansari2018}, leads to 
\begin{equation}
\label{schmidt_expansion}
\begin{aligned}
    \hat{a}^{\dagger}(\omega_{s}) &= \sum_{k=1}^{N} u_{k}^{(s)*}(\omega_{s})\, \hat{A}_{k}^{(s)\dagger}(\omega_{s})\,,\\
    \hat{b}^{\dagger}(\omega_{i}) &= \sum_{k=1}^{N} v_{k}^{(i)*}(\omega_{i})\, \hat{B}_{k}^{(i)\dagger}(\omega_{i}).  
\end{aligned}
\end{equation}
We substitute these expressions together with the Schmidt decomposition of the JSA into Eq.\eqref{SPDC_Hamiltonian}, and using the orthogonality of the Schmidt modes:
\begin{equation}
\begin{aligned}
    &\int u_k^{(s)*}(\omega_{s})\,u_l^{(s)}(\omega_{s})\,d\omega_{s}&=\delta_{kl}, \\
    &\int v_k^{(i)*}(\omega_{i})\,v_l^{(i)}(\omega_{i})\,d\omega_{i}&=\delta_{kl},
\end{aligned}
\end{equation}
we get a diagonalized interaction Hamiltonian in the Schmidt basis
\begin{equation}
     \hat{H}_{I} = i\hbar\,\left[\sum_{k=1}^{N}r_{k}\left(\hat{A}_{k}^{(s)\dagger}(\omega_{s})\hat{B}_{k}^{(i)\dagger}(\omega_{i})-\hat{A}^{(s)}_{k}(\omega_{s})\hat{B}^{(i)}_{k}(\omega_{i})\right)\right]\,.
     \quad
     \label{MM_hamiltonian}
\end{equation}
Here, $r_{k} = r \lambda_{k}$ is the effective squeezing parameter associated with the k\textsuperscript{th} Schmidt mode. The corresponding multimode squeezing operator is given by
\begin{equation}
     \hat{S}(r_{k}) = \exp\,\left[\sum_{k=1}^{N}r_{k}\left(\hat{A}_{k}^{(s)\dagger}\hat{B}_{k}^{(i)\dagger}-\hat{A}^{(s)}_{k}\hat{B}^{(i)}_{k}\right)\right]\, .
\end{equation}

Throughout this work, we assume that the two OPAs possess identical Schmidt-mode structures, corresponding to the same Schmidt basis and identical mode-dependent squeezing parameters. Furthermore, the phase accumulated between the two amplifiers is assumed to be independent of the Schmidt mode, such that every mode acquires the same phase during propagation through the interferometer. Under these assumptions, the evolution operator of the complete SU(1,1) interferometer (Figure\,\ref{fig1_SU11_block_dia}-A) is given by 
\begin{equation}
    \hat{U}_{SU(1,1)} = \hat{S}_{2}\, \hat{P}(\varphi)\, \hat{S}_{1}\,.
    \label{SU_operator}
\end{equation}
Where $\hat{S}_{1/2}$ denotes the squeezing operator of the first and the second OPA, and $\hat{P}(\varphi)$ represents the intermediate phase-shift operator placed in the signal arm, since each of these operators became diagonal independently in the Schmidt basis, 
\begin{equation}
\hat{S}_1=\bigotimes_k \hat{S}_{1,k},
\qquad
\hat{P}(\varphi)=\bigotimes_k \hat{P}_k(\varphi),
\qquad
\hat{S}_2=\bigotimes_k \hat{S}_{2,k}.
\end{equation}
The interferometer evolution operator given by Eq. \eqref{SU_operator} can be expressed as 
\begin{equation}
\begin{aligned}
\hat{U}_{\mathrm{SU}(1,1)}
&=
\left(\bigotimes_{k=1}^{N} \hat{S}_{2,k}\right)
\left(\bigotimes_{k=1}^{N} \hat{P}_k(\varphi)\right)
\left(\bigotimes_{k=1}^{N} \hat{S}_{1,k}\right)\\
&=
\bigotimes_{k=1}^{N}
\left(
\hat{S}_{2,k}\,
\hat{P}_k(\varphi)\,
\hat{S}_{1,k}
\right).
\end{aligned}
\end{equation}
Defining $\hat{U_{k}} = \hat{S}_{2,k}\, \hat{P}_{k}(\varphi)\, \hat{S}_{1,k}$, the SU(1,1) interferometric transformation operator for the k\textsuperscript{th} mode, one can finally write the overall transformation operator of  multimode SU(1,1) transformation as
\begin{equation}
\hat{U}_{\mathrm{SU}(1,1)}
=
\bigotimes_{k=1}^{N}
\hat{U}_k.
\label{SU_MM_trans}
\end{equation}

Therefore, under the assumption stated above, Eq. \eqref{SU_MM_trans} shows that a multimode SU(1,1) interferometer is mathematically equivalent to a tensor product of independent single-mode SU(1,1) interferometers when represented in the Schmidt basis. Each Schmidt-mode pair evolves independently with its own squeezing parameter ($r_{k}$). At the same time,e the multimode nature of the interferometer arises solely from the distribution of the total squeezing resource among the Schmidt modes and the way their individual contributions are combined through the measurement process.

\subsubsection{Input State in Schmidt Basis}
The input state considered throughout this work is $\ket{\psi_{in}} = \ket{\alpha_{f}}_{s} \otimes \ket{0}_{i}$, where the signal arm receives a coherent state in a normalized spectral mode $f(\omega)$, with mean photon number $|\alpha|^{2}$, and the idler arm is kept at vacuum. A broadband annihilation operator generates the coherent state \cite{20_PhysRevX.5.041017,20_Ansari2018}
\begin{equation}
\hat{a}_{f} = 
\int d\omega\, f(\omega)\,\hat{a}(\omega).
\end{equation}
Since the $\{u_{k}^{(s)}(\omega)\}$ form a complete orthonormal basis, the spectral mode function $f(\omega)$ can be expanded as-
\begin{equation}
\label{coherent_coeff}
    f(\omega) = \sum_{k=1}^{N} c_{k}\,u_{k}^{(s)}(\omega)
    \qquad
    \mathrm{where}
    \qquad
    c_{k} = \int d\omega\, u_{k}^{(s)*}(\omega) \,f(\omega)\,.
\end{equation}
As $f(\omega)$ is normalized, the coefficients $c_{k}$ must satisfy $\sum_{k}|c_{k}|^{2} = 1$. The creation operator $\hat{a}_{f}^{\dagger}$ of the coherent mode $f(\omega)$ can be written in terms of the operators $\hat{A}_{k}^{(s)\dagger}(\omega_{s})$
\begin{equation}
\hat{a}_f^{\dagger}
=
\int d\omega\, f^*(\omega)\hat{a}^{\dagger}(\omega)
=
\sum_{k=1}^{N} c_k^*
\int d\omega\, u_k^{(s)*}(\omega)\hat{a}^{\dagger}(\omega)
=
\sum_{k=1}^{N} c_k^* \hat{A}^{(s)\dagger}_k.
\end{equation}
Thus, the argument of the exponential in the displacement operator of the coherent state can be written as 
\begin{equation}
\alpha \hat{a}_f^\dagger-\alpha^*\hat{a}_f
=
\sum_{k=1}^{N}\left(\alpha c_k^* \hat{A}_k^{(s)\dagger}(\omega_{s})-\alpha^* c_k \hat{A}^{(s)}_k(\omega_{s})\right).
\end{equation}
Since the displacement operators associated with different Schmidt modes commute, the input state factorizes according to 
\begin{equation}
\ket{\psi_{\mathrm{in}}} = \bigotimes_{k=1}^{N} \left(
\ket{\alpha c_k^{*}}_{A_k^{(s)}(\omega_{s})}  \otimes \ket{0}_{B^{(i)}_{k}{(\omega_{i})}}\right).
\label{input_decomposition}
\end{equation}
The coefficient $c_{k}$ given by Eq.\,\eqref{coherent_coeff} describes the overlap between the spectral mode $u^{(s)}_{k}(\omega_{s})$ of the OPA with the coherent input mode $f(\omega)$. Equation \eqref{input_decomposition} together with Eq. \eqref{SU_MM_trans} shows that a multimode SU(1,1) interferometer probed by a $\ket{\psi_{in}} = \ket{\alpha_{f}}_{s} \otimes \ket{0}_{i}$  input can be interpreted as a collection of independent single-mode interferometers, each associated with the Schmidt-mode pair $(u_{k}^{(s)}(\omega_{s}),(v_{k}^{(i)}(\omega_{i}))$. Each interferometer is characterized by a squeezing parameter $r_{k} = r\lambda_{k}$ given by Eq.\,\eqref{MM_hamiltonian} and probed with a input state $\ket{\psi_{\mathrm{k, in}}} = \ket{\alpha c_k^{*}}_{A_k^{(s)}(\omega_{s})} \otimes \ket{0}_{B^{(i)}_{k}{(\omega_{i})}}$. As both the total interferometric transformation and the input state can be diagonalized in the Schmidt basis of the OPA, the multimode nature and its effects will manifest themselves in how the total squeezing strength is distributed among the different $k$ modes and how they are combined in the measurement process.  

\subsubsection{Mode-Resolved Input-Output Relation}
Each Schmidt-mode pair evolves independently through the SU(1,1) interferometer and therefore obeys the same Bogoliubov transformation as a conventional single-mode interferometer, with the multimode nature entering solely through the mode-dependent squeezing parameter. The transformation corresponding to the $k$\textsuperscript{th} Schmidt mode of the first OPA reads \cite{11_Ou2020} 
\begin{equation}
\hat{A}_{k,1}^{(s)} = G_{1,k}\hat{A}^{(s)}_{k,in} + g_{1,k}\hat{B}_{k,in}^{(i)\dagger}
\quad
\hat{B}_{k,1}^{(i)} = G_{1,k}\hat{B}^{(i)}_{k,in} + g_{1,k}\hat{A}_{k,in}^{(s)\dagger}.
\end{equation}
Here, 
$ G_{1,k}=\cosh{r_{1,k}}$ and $g_{1,k}=\sinh{r_{1,k}} $  are the gain coefficients for the $k^{\mathrm{th}}$ mode. After propagation through the phase shift section and  subsequent amplification in the second OPA, the output operators  are given, in terms of the input operator, by 
\begin{equation}
\begin{aligned}
\hat{A}^{(s)}_{k,\mathrm{out}}(\omega_{s})
&=
U_k(\varphi)\hat{A}^{(s)}_{k,\mathrm{in}}(\omega_{s})
+
V_k(\varphi)\hat{B}_{k,\mathrm{in}}^{(i)\dagger}(\omega_{i})\ ,\\
\hat{B}^{(i)}_{k,\mathrm{out}}(\omega_{i})
&=
\widetilde{U}_k(\varphi)\hat{B}^{(i)}_{k,\mathrm{in}}(\omega_{i})
+
\widetilde{V}_k(\varphi)\hat{A}_{k,\mathrm{in}}^{(s)\dagger}(\omega_{s})\ ,
\end{aligned}
\end{equation}
where 
\begin{equation}
\label{Trans_coefficient}
\begin{aligned}
U_k(\varphi)
&=
G_{2,k}G_{1,k}e^{i\varphi}
+
g_{2,k}g_{1,k}\ ,\\
V_k(\varphi)
&=
G_{2,k}g_{1,k}e^{i\varphi}
+
g_{2,k}G_{1,k},\\
\widetilde{U}_k(\varphi)
&=
G_{2,k}G_{1,k}
+
g_{2,k}g_{1,k}e^{-i\varphi}\ ,\\
\widetilde{V}_k(\varphi)
&=
G_{2,k}g_{1,k}
+
g_{2,k}G_{1,k}e^{-i\varphi}.
\end{aligned}
\end{equation}
As discussed in Section \ref{Multimode_SU(1,1)_Interferometer}, the two OPAs are identical and share the same mode-dependent squeezing parameter. This implies that $G_{1,k}=G_{2,k}\equiv G_{k}$ and $g_{1,k}=g_{2,k}\equiv g_{k}$. These mode-resolved input-output relations constitute the fundamental building block for the analysis that follows. Since every Schmidt mode evolves independently according to the same SU(1,1) transformation, all measurement observables may be constructed by combining the contributions from the individual modes. Consequently, the distinction between single-mode and multimode interferometry arises not from the evolution itself, but from the manner in which the different modes contribute to the measured signal.

\subsection{Phase Sensitivity}
\label{sensitivity}
The ultimate performance of an interferometer is characterized by its phase sensitivity, which quantifies the smallest measurable phase shift for a given measurement strategy. The central objective of this work is to investigate how the multimode structure of the OPA influences the achievable phase sensitivity. Although the evolution of each mode is independent, the measured sensitivity generally depends on how the detection process combines the contributions from individual modes. Consequently, the multimode nature of the interferometer can degrade sensitivity relative to the ideal single-mode case.

Within the error-propagation formalism \cite{19_Ghosh2026}, the phase sensitivity associated with an arbitrary measurement operator $\hat{O}$ is given by
\begin{equation}
\Delta\varphi
=
\frac{\Delta\hat{O}}
{\left|\dfrac{\partial\langle\hat{O}\rangle}{\partial\varphi}\right|}.
\label{sensitivity_formula}
\end{equation}
where $\langle \hat{O}\rangle$ and $\Delta\hat{O}$ denote the expectation value and standard deviation of the measurement observable, respectively.
The chosen detection scheme determines the specific form of the measurement operator.

In this work, we consider two experimentally relevant measurement strategies: number and balanced homodyne detection. We are interested in quantifying the interferometer's sensitivity degradation due to its multimode nature. We will only consider the total output photon number in the idler mode, i.e., $\hat{O} =\sum_{k} \hat{B}_{{k, out}}^{\dagger(i)}\, \hat{B}_{k, out}^{(i)}$. Similarly, balanced homodyne detection at the signal output port with the same input state provides the best sensitivity in the single-mode case. Therefore, for the multimode case, our choice of operator is the output quadrature $\hat{O} = \hat{Y}^{(s)}_{g,{\mathrm{out}}}(\theta)$ at the signal output port. In this work, the local-oscillator reference phase is set to $\theta=0$, corresponding to the optimal phase sensitivity, while $g(\omega)$ denotes the spectral mode profile of the local oscillator. 

The phase sensitivities for these two measurement schemes are derived separately in the following section and subsequently compared with the single-mode sensitivity to quantify the influence of the multimode structure on the interferometric performance.

\section{Sensitivity Analysis}
\label{sen_exp}
The phase sensitivity of the multimode SU(1,1) interferometer is analyzed for the two measurement schemes introduced in Section \ref{sensitivity}. The observables related to those measurements are $\hat{O} =\sum_{k} \hat{B}_{{k, out}}^{\dagger(i)}\,\hat{B}_{k, out}^{(i)} = \sum_{k}\hat{N}_{{k, out}}^{(i)} $ (total photon number at the idler output) for number detection and $\hat{O} = \hat{Y}^{(s)}_{g,{\mathrm{out}}}(\theta)$ (quadrature measurement at the signal output) for homodyne detection.

\subsection{Number Detection}
\label{num_theory}
Since both the multimode interferometer and the input coherent state factorize into independent Schmidt modes, the phase sensitivity can be obtained by considering the contributions of each mode. It is well known in the literature that for a single-mode interferometer with number detection, both the mean photon number and the photon-number fluctuation vanish at the dark fringe $(\varphi = \pi)$ at the idler output \cite{11_Ou2020}. Following the standard single-mode analysis, we therefore expand the mean (signal) and the variance (noise) around the operating point $\varphi = \pi + \epsilon$ with $\epsilon \ll \pi$. To the leading order, the mean and the variance at the idler output of the $k$\textsuperscript{th} Schmidt modes are 
\begin{equation}
\langle \hat{N}_{k,out}^{(i)} \rangle_{\varphi = \pi + \epsilon}
\simeq
G_k^2 g_k^2 \left(1+|c_{k}\,\alpha|^2\right)\epsilon^2,
\end{equation}
and
\begin{equation}
\left(\Delta \hat{N}_{k,out}^{(i)}\right)^2_{\varphi = \pi + \epsilon}
\simeq
G_k^2 g_k^2 \left(1+|c_{k}\,\alpha| ^2\right)\epsilon^2.
\end{equation}
Since different Schmidt modes evolve independently, the total output signal is simply the sum of the signals corresponding to the individual modes
\begin{equation}
    \langle\hat{N}_{out}^{(i)}\rangle_{\varphi = \pi + \epsilon} = \sum_{k} \langle \hat{N}_{k,out}^{(i)} \rangle_{\varphi = \pi + \epsilon} =\,\epsilon^{2} \sum_{k}\, G_{k}^{2}\, g_{k}^{2} \, \left( 1 + |c_{k} \alpha|^{2}\right)\ .
\end{equation}
Similarly, the total variance is 
\begin{equation}
    \left( \Delta \hat{N}_{out}^{(i)}\right)^{2} = \sum_{k} \left( \Delta \hat{N}_{k,out}^{(i)}\right)^{2}\ .
\end{equation}
Here, the covariance terms vanish because the photo-number fluctuations associated with different Schmidt modes are uncorrelated ($\mathrm{Cov}(\hat{N}_{l}$, $\hat{N}_{m}) = 0, l \neq m )$. So the total photon number variance at the idler output becomes 
\begin{equation}
    \left(\Delta \hat{N}_{out}^{(i)}\right)^{2}_{\varphi = \pi + \epsilon} = \sum_{k}  \left(\Delta \hat{N}_{k,out}^{(i)}\right)^{2}_{\varphi = \pi + \epsilon} =\,\epsilon^{2} \sum_{k}\, G_{k}^{2}\, g_{k}^{2} \, \left( 1 + |c_{k} \alpha|^{2}\right)\ .
\end{equation}
Substituting the mean and variance in the error-propagation formula, Eq. \eqref{sensitivity_formula}, yields the following sensitivity at the dark fringe $(\varphi = \pi)$ with number detection for a multimode interferometer 
\begin{equation}
    \left(\Delta \varphi\right)_{MM}^{\hat{N}_{out}^{(i)}} = \frac{1}{2\,\sqrt{\sum_{k}\, G_{k}^{2}\, g_{k}^{2} \, \left( 1 + |c_{k} \alpha|^{2}\right)}}\ .
    \label{mm_sen_number}
\end{equation}
The corresponding sensitivity expression for a single-mode interferometer is 
\begin{equation}
    \left(\Delta \varphi \right)_{\mathrm{SM}}^{\hat{N}_{\mathrm{out}}^{\mathrm{(i)}}} = \frac{1}{2\,\sqrt{G^{2}g^{2}(1 + |\alpha|^{2})}}\ .
    \label{sm_sen_number}
\end{equation}
It turns out, by comparing Eqs. \eqref{mm_sen_number} and \eqref{sm_sen_number} that $\left(\Delta \varphi\right)_{MM}^{\hat{N}_{out}^{(i)}} > \left(\Delta \varphi \right)_{\mathrm{SM}}^{\hat{N}_{\mathrm{out}}^{\mathrm{(i)}}}$ for $k>1$, i.e., the multimode nature of the interferometer always degrades the sensitivity with respect to the single-mode case. Considering the same squeezing parameter $(r= \mathrm{constant})$ for both cases, the sensitivity degradation in the multimode case is solely due to the distribution of the fixed total squeezing strength among the different Schmidt modes. Since the phase-sensitive response scales nonlinearly with gain, several weaker amplifiers cannot produce the same signal as a single strong amplifier using the same total resource. Since the denominator of Eq. \eqref{mm_sen_number} is maximized by concentrating the nonlinear gain into the dominant Schmidt mode, the best, i. e., minimum, phase sensitivity in the multimode case is achieved when the two OPAs operate as close to the single-mode regime as possible. Furthermore, for a fixed coherent input power, the coherent field should be mode-matched to the highest-gain Schmidt mode so that the largest part of the input light experiences the strongest parametric amplification.      
\subsection{Homodyne Detection}
\label{hd_theory}
In balanced homodyne detection, the measured quadrature is determined by the LO's spectral shape. We consider a local oscillator occupying a normalized spectral mode $g(\omega)$. Expanding this mode in the Schmidt basis and writing the annihilation operator corresponding to mode $g(\omega)$ in terms of the Schmidt mode annihilation operator
\begin{equation}
\hat{A}^{(s)}_{g,out} =  \sum_{k=1}^{N} d_{k}^{*} \hat{A}^{(s)}_{k,out}\ ,
\end{equation}
where
\begin{equation}
d_{k} = \int\, d\omega\, u^{*}_{k}(\omega) g(\omega)\ ,
\end{equation}
and where the coefficients $d_{k}$ satisfy $\sum_{k=1}^{N}|d_{k}|^{2} = 1 $. Now, the measured quadrature operator is
\begin{equation}
  \hat{Y}_{g,\mathrm{out}}^{(s)} = i \left( \hat{A}_{g,out}^{\dagger(s)} - \hat{A}^{(s)}_{g,out}\right)\ .
\end{equation}
For a balanced interferometer, operating at the dark fringe ($\varphi = \pi$), the phase derivative of the quadrature mean is
\begin{equation}
\left|
\frac{\partial}{\partial \varphi}
\left\langle
\hat{Y}_{g,\mathrm{out}}^{(s)}
\right\rangle
\right|_{\varphi=\pi}
=
\,2\,\mathrm{Re} \left[\alpha\sum_{k}G_{k}^{2}\,d_{k}^{*}\,c_{k}^{*}
\right]\ ,
\end{equation}

and its variance is $\left(\Delta\hat{Y}_{g,\mathrm{out}}^{(s)}\right)^{2} = 1\ $. Here, we have used the independence of the Schmidt modes, i. e., $\left\langle
\Delta\hat{Y}_{i,\mathrm{out}}^{\mathrm{(s)}}
\Delta\hat{Y}_{j,\mathrm{out}}^{\mathrm{(s)}}
\right\rangle
=
\delta_{ij}
\left(\Delta Y_{i,\mathrm{out}}^{\mathrm{(s)}}\right)^2$.

Finally, the phase sensitivity (Eq. \eqref{sensitivity_formula}) for the multimode interferometer with homodyne detection at the signal output is given by
\begin{equation}
    \left(\Delta \varphi\right)_{\mathrm{MM}}^{\hat{Y}_{out}^{\mathrm{(s)}}} = \frac{1}{2\, \mathrm{Re}\,\left[ \alpha\, \sum_{k}  G_{k}^{2}\,d_{k}^{*}\, c_{k}^{*}\right]}\ .
    \label{MM_hd}
\end{equation}
For comparison, the phase sensitivity of an ideal single-mode SU(1,1) interferometer with balanced homodyne detection is \cite{14_Jana2026} 
\begin{equation}
    \left(\Delta \varphi\right)_{\mathrm{SM}}^{\hat{Y}_{out}^{(\mathrm{s})}} = \frac{1}{2\,\mathrm{Re}\,\left[\alpha\,  G^{2}\right]}\ .
    \label{SM_hd}
\end{equation}
Similarly to the case of number measurement, Eqs. \eqref{MM_hd} and \eqref{SM_hd} show that $\left(\Delta \varphi\right)_{\mathrm{MM}}^{\hat{Y}_{out}^{(s)}} \geq \left(\Delta \varphi \right)_{\mathrm{SM}}^{\hat{Y}_{\mathrm{out}}^{\mathrm{(s)}}}$, i.e., the multimode nature of the interferometer can only degrade its sensitivity. The equality holds only when $k = 1$ and both the LO and the input coherent state are perfectly mode-matched with the first Schmidt mode. Since the phase sensitivity depends explicitly on the local-oscillator mode through the overlap term $\sum_{k}G_{k}^{2}\,d_{k}^{*}\,c_{k}^{*}$, it can be minimized by optimizing the modal profile of the local oscillator. The optimal local oscillator is obtained as 
\begin{equation}
    d_{k}^{\mathrm{opt}} = \frac{c_{k}^{*}G_{k}^{2}}{\sqrt{\sum_{j}{|c_{j}|^{2}G_{j}^{4}}}}\ ,
    \label{opt_lo}
\end{equation}
yielding the minimum achievable phase sensitivity in the multimode case
\begin{equation}
    \left(\Delta \varphi\right)_{MM}^{\mathrm{opt}} = \frac{1}{2\,\alpha\,\sqrt{ \sum_{k} G_{k}^{4}\, |c_{k}|^{2}}}\ .
    \label{MM_hd_opt}
\end{equation}

Unlike number detection, the sensitivity degradation in homodyne detection may originate from two distinct physical mechanisms. The first one is the fragmentation of the available squeezing strength among multiple Schmidt modes. For a fixed total nonlinear resource, each Schmidt mode experiences a smaller parametric gain than the corresponding single-mode interferometer. Even when the coherent probe is perfectly mode-matched to the dominant Schmidt mode, the minimum attainable sensitivity is worse than the single-mode case as $G_{1} < G$. The second mechanism is unique to homodyne detection. Balanced homodyne detection measures the optical field quadrature; therefore, the measurement outcome depends on the coherent overlap between the signal and LO fields through $\sum_{k}G_{k}^{2}\,d_{k}^{*}\,c_{k}^{*}$. Any mismatch between the modal distributions of the coherent probe and the local oscillator in the Schmidt basis reduces the detected field quadrature value and, consequently, degrades the phase sensitivity. Although the optimal local oscillator (given by Eq. \eqref{opt_lo}) maximizes this overlap, it cannot compensate for the sensitivity degradation caused by the nonlinear gain distribution. 

These results highlight a fundamental distinction between the two detection schemes. In photon-number detection, the multimode penalty originates solely from the redistribution of the finite nonlinear resource among independent Schmidt modes. Homodyne detection inherits this unavoidable resource-fragmentation penalty but additionally suffers from a coherence penalty associated with the modal overlap between the signal and the LO. Consequently, homodyne detection is intrinsically more sensitive to the multimode structure of the parametric process than photon-number detection.

\section{Numerical Simulation of PPLN Waveguides}
The multimodal structure of a waveguide-based OPA is governed by the underlying parametric process and is completely characterized by the Schmidt decomposition of its JSA. While the theoretical framework developed in section \ref{Theory} is independent of a particular waveguide design, evaluating the phase sensitivity of a realistic SU(1,1) interferometer requires the Schmidt-mode distribution of the corresponding PDC source.

In this work, we consider a PP-TFLN waveguide for the representative realization of the OPA. The PDC process is numerically simulated to obtain the JSA of the parametrically generated light, on which the Schmidt decomposition is performed to extract the Schmidt coefficients $(\lambda_{k})$. As discussed in Section\,\ref{Multimode_SU(1,1)_Interferometer}, these coefficients establish the direct connection between the OPA and the multimode phase-sensitivity analysis presented in this work.
\newline
\begin{strip}
\centering
\includegraphics[width=1.0\linewidth]{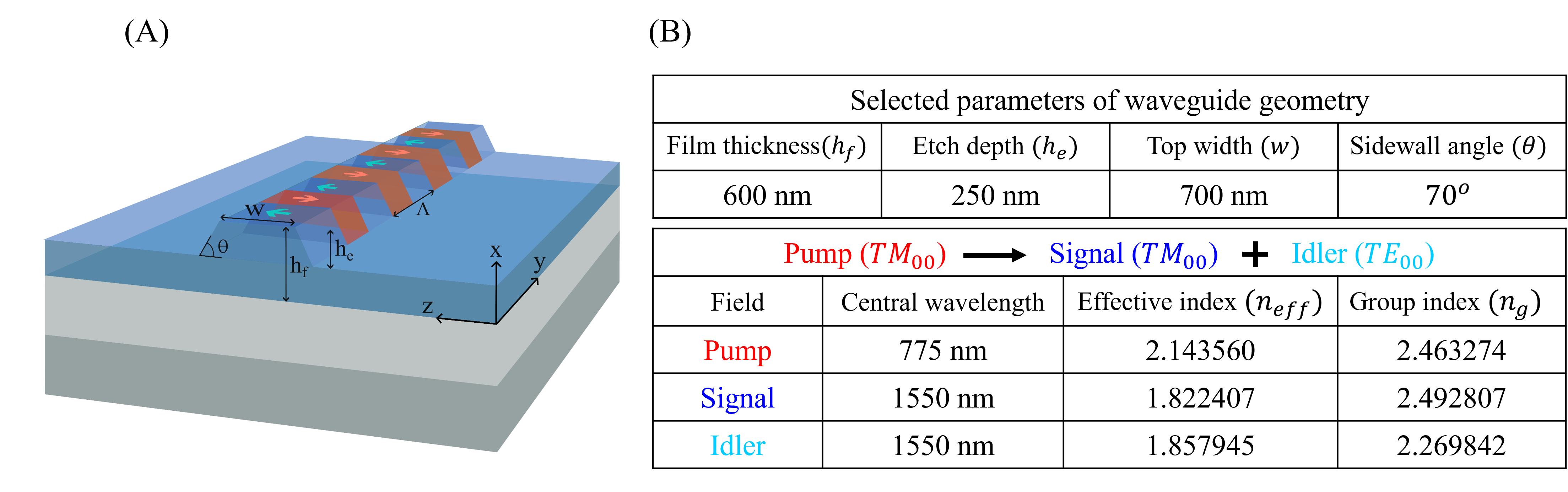}
\captionof{figure}{Waveguide configuration, design parameters, and simulated modal properties of the PP-TFLN waveguide employed for Type-II SPDC. (A) Schematic of the shallow-etched PP-TFLN waveguide implemented on an $x$-cut $5\%$ MgO-doped TFLN platform. Light propagates along the $y$-axis (normal to the optic axis of LN). (B) Waveguide geometric parameters (top table) are selected to satisfy the asymmetric group-velocity-matching (GVM) criterion, $n_g^s \approx n_g^p $, while the corresponding group indices (bottom table) are obtained from eigenmode simulations for the pump, signal, and idler modes. The corresponding effective refractive indices yield a required first-order QPM poling period of $\Lambda \approx 2.55~\mu\mathrm{m}$.}
\label{fig2_Waveguide_geometry}

\centering
\includegraphics[width=1.0\linewidth]{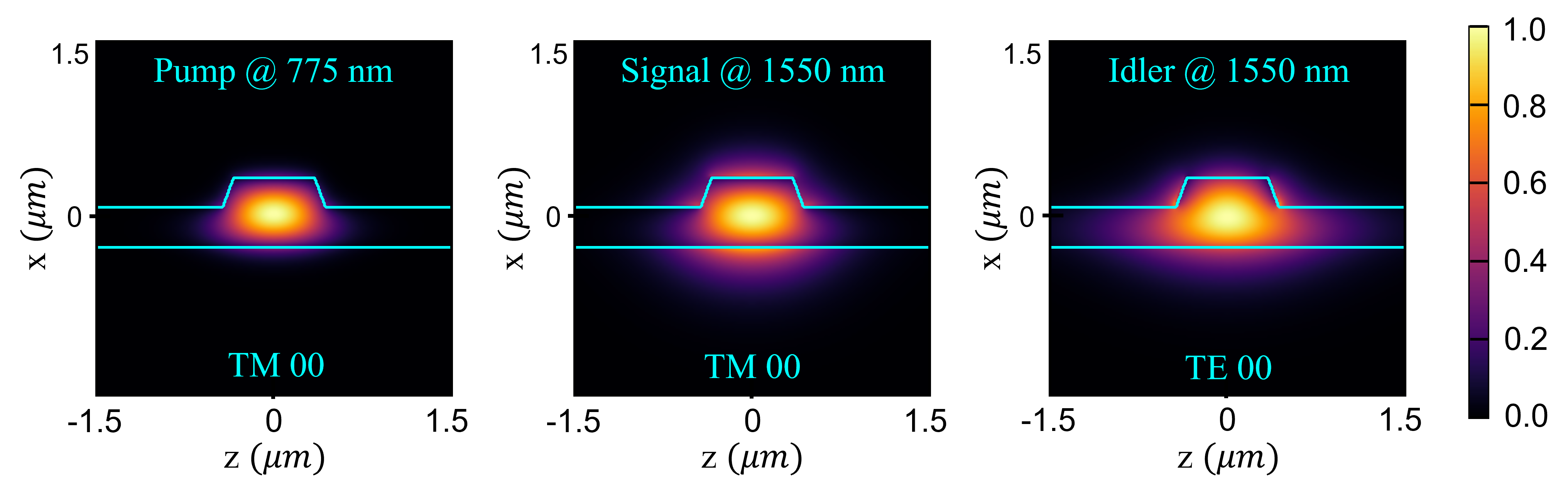}
\captionof{figure}{Normalized spatial profiles of the absolute electric-field magnitude for the pump, signal, and idler fundamental guided modes supported by the designed waveguide, obtained from Lumerical eigenmode simulations.}
\label{fig3_Waveguide_mode}
\end{strip}

\subsection{Waveguide Design}
\label{waveguide_design}
The OPA is realized using a PP-TFLN  rib waveguide as shown in Figure\,\ref{fig2_Waveguide_geometry}-A, designed for a degenerate type-II PDC in which $775 \,\mathrm{nm}$ TM-polarized pump light is converted into a pair of TM-polarized signal and TE-polarized idler, both centered at $1550 \,\mathrm{nm}$. An x-cut lithium niobate substrate with light propagating along the $y\,\mathrm{-axis}$ is employed in the simulation. The simulated waveguide geometry and the corresponding design parameters are summarized in Figure\,\ref{fig2_Waveguide_geometry}-B. The waveguide dimensions are designed to support single-spatial-mode operation at the signal and idler wavelengths while remaining only weakly multimode at the pump wavelength, thereby ensuring that all three interacting fields propagate predominantly in their respective fundamental guided modes. The corresponding fundamental mode profiles, obtained using the finite-element mode solver in Lumerical, are shown in Figure\,\ref{fig3_Waveguide_mode}. The simulated effective refractive indices are subsequently used to evaluate the waveguide's phase-matching characteristics required for the numerical simulation of the parametric conversion. The poling period is set for degenerate signal-and-idler generation at $1550 \,\mathrm{nm}$. The chosen waveguide dimensions (Figure\,\ref{fig2_Waveguide_geometry}-B) ensure generation of photon pairs with high spectral purity through asymmetric group-velocity matching, thereby minimizing the spectral correlation between the generated signal and idler photons. The resulting PDC simulation yields the JSA of the generated photon pair, on which the Schmidt decomposition is performed to obtain the modal structure of the OPA.
\newpage
\begin{strip}
\centering
\includegraphics[width=1.0\linewidth]{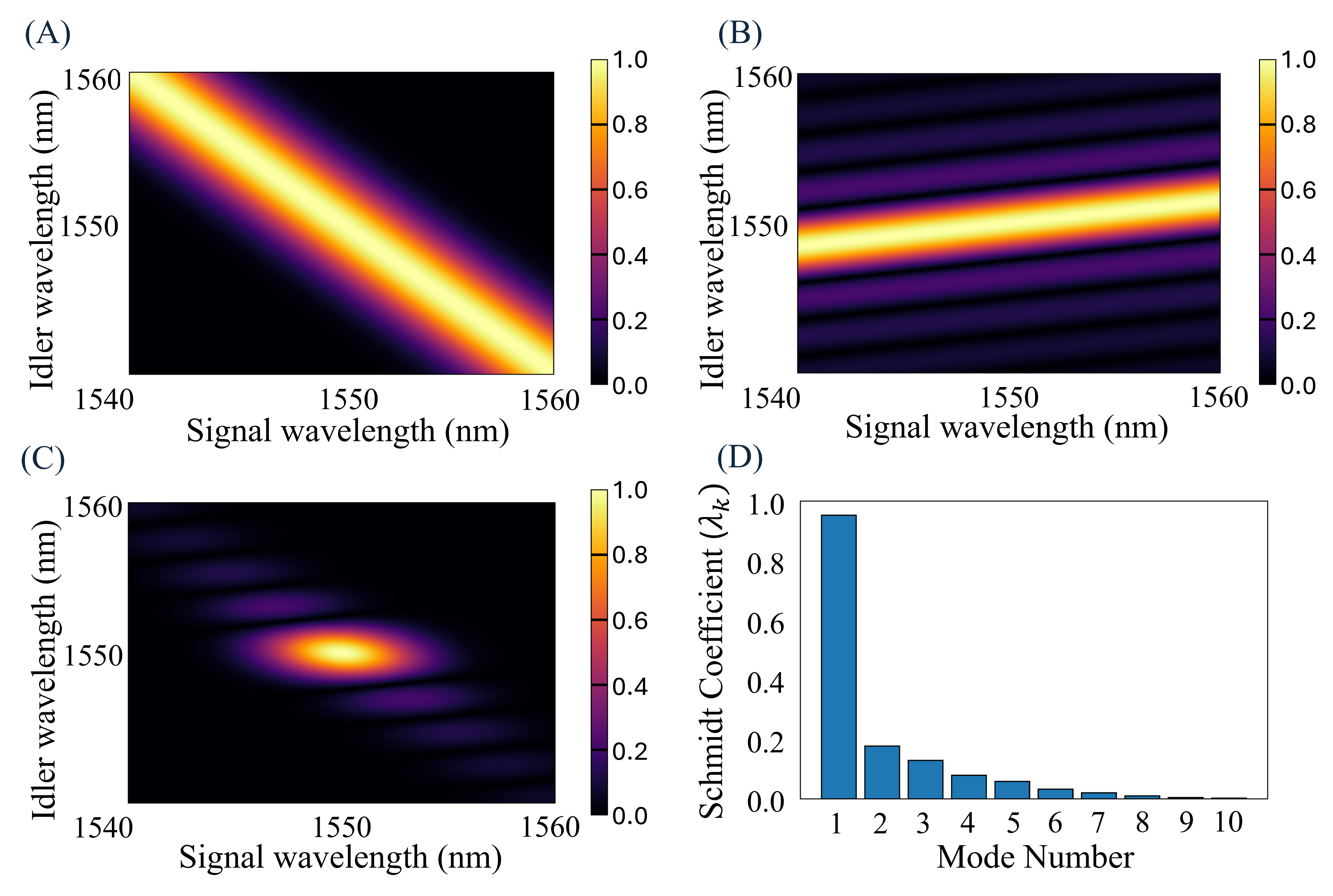}
\captionof{figure}{Spectral properties of the PDC process. (A) Gaussian pump spectrum centred as \(775~\mathrm{nm}\) pump with bandwidth of \(2~\mathrm{nm}\) at FWHM. (B) Numerically simulated phase-matching spectrum of the designed PP-TFLN waveguide. (C) Absolute JSA of the simulated PDC process. (D) Schmidt coefficient distribution of the simulated JSA.}
\label{fig4_Schmidt_mode}
\vspace{12pt}
\centering
\includegraphics[width=1.0\linewidth]{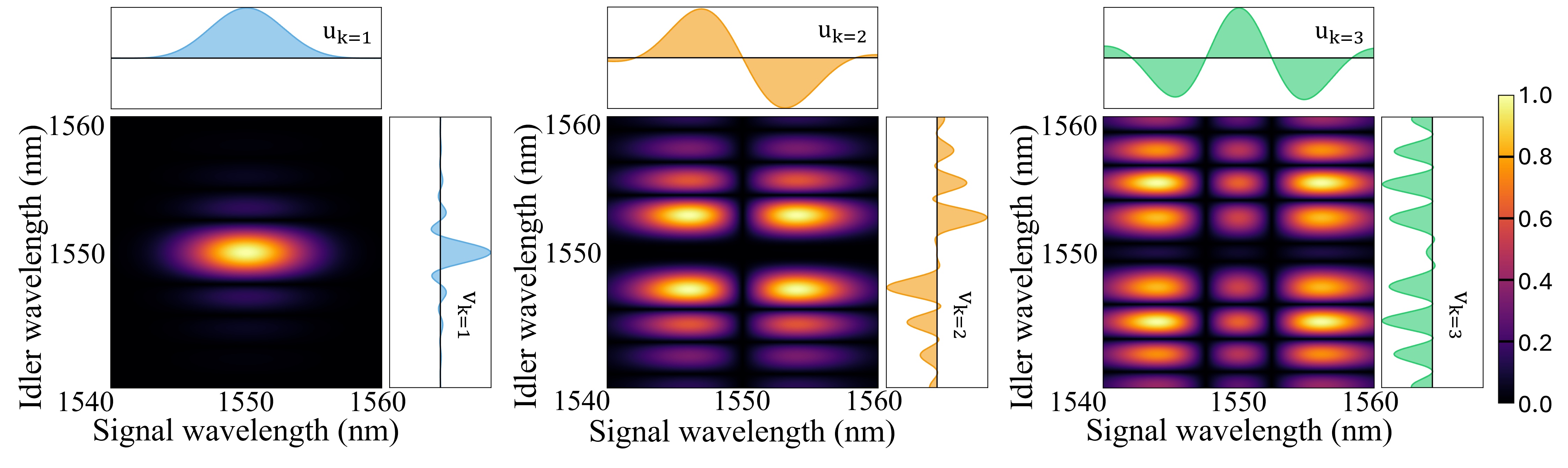}
\captionof{figure}{The first three Schmidt modes of the JSA. The central panels show the absolute value of the joint spectral amplitude associated with each Schmidt mode, while the upper and right panels show the corresponding real parts of the signal and idler fields, respectively. The mode profiles are obtained for a pump FWHM of $2$ nm and a waveguide length of $5$ mm, yielding a spectral purity of $89.7\%$.}
\label{fig5_Schmidt_mode_profiles}
\end{strip}

\subsection{Spectral Mode Analysis}
The spectral properties of the generated photon pairs are characterized through the JSA, whose Schmidt decomposition determines the multimode structure of the OPA, as shown in Figure\,\ref{fig4_Schmidt_mode}. We consider a Gaussian pump spectrum centered at $775\, \mathrm{nm}$ with a spectral bandwidth of $2\,\mathrm{nm}$ at full-width-half-maximum (FWHM), leading to the pump spectral distribution of Figure\,\ref{fig4_Schmidt_mode}-A. The waveguide geometry described in section \ref{waveguide_design} corresponds to a phase matching angle, i. e., the angle made by the tangent of the phase matching function with the horizontal axis of Figure\,\ref{fig4_Schmidt_mode}-B, of approximately $9^{\circ}$. In the JSA simulation, the waveguide length is kept fixed at $5\,\mathrm{mm}$. The combination of the pump spectrum of Figure\,\ref{fig4_Schmidt_mode}-A and the phase-matching spectrum of Figure\,\ref{fig4_Schmidt_mode}-B leads to the JSA, whose absolute value is shown in Figure\,\ref{fig4_Schmidt_mode}-C. The first ten Schmidt coefficients are presented in Figure\,\ref{fig4_Schmidt_mode}-D. The corresponding spectral profiles of the first three Schmidt modes are shown in Figure\,\ref{fig5_Schmidt_mode_profiles}, illustrating the signal and idler Schmidt modes together with the associated absolute joint spectral amplitude distribution. The Schmidt decomposition yields a spectral purity $P = \sum_{k}\lambda_{k}^{4}$  of approximately $89.7\,\% $, demonstrating that the designed waveguide generates a nearly factorizable two-photon state with a small residual multimode component. Although TFLN waveguide designs with higher spectral purity have been reported \cite{Zhao2020}, the proposed design offers sufficient spectral purity to serve as a realistic and representative platform for the present sensitivity analysis of multimode SU(1,1) interferometers. 

\begin{figure}[h]
    \centering
    \includegraphics[width=1\linewidth]{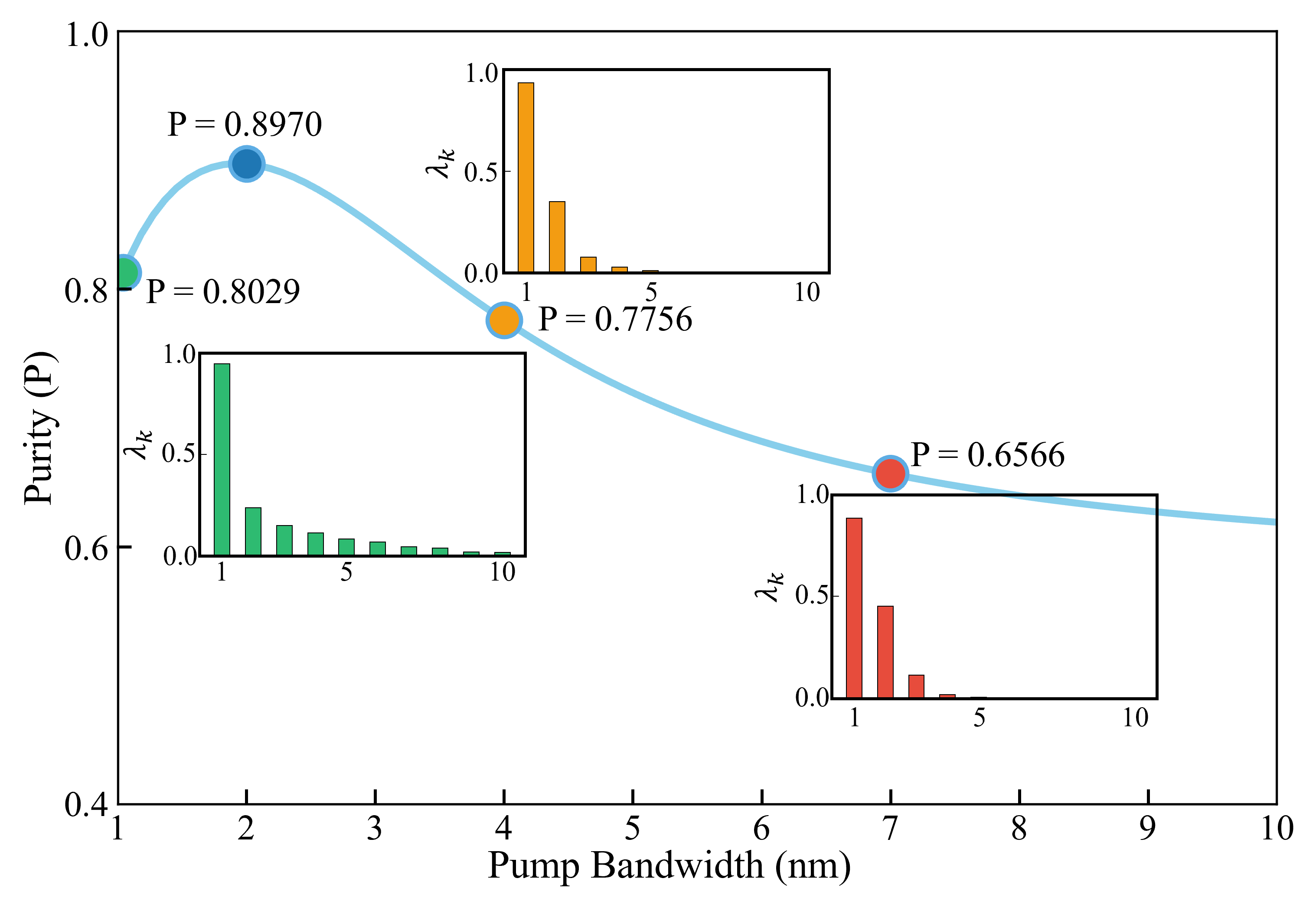}  \caption{Purity of the generated photon pairs from the simulated PDC process as a function of the pump bandwidth. The inset shows the Schmidt coefficient distribution for three representative purity values, illustrating the increasing contribution of the higher-order Schmidt modes as the purity decreases.}  
    \label{fig_purity}
\end{figure}
\vspace{1pt}

In the following, to quantify the impact of this residual multimode structure on the phase sensitivity of the SU(1,1) interferometer, the Schmidt coefficients are calculated for a pump FWHM varying from $1$ to $10\,\mathrm{nm}$, while the waveguide length is kept fixed at $5\, \mathrm{mm}$. This keeps the nonlinear interaction strength, and hence the total parametric gain, unchanged throughout the analysis. Changing the pump bandwidth, therefore, modifies only the distribution of the Schmidt coefficients, redistributing the fixed squeezing resource among the Schmidt modes. The Schmidt coefficients are then used in the sensitivity expression derived in Section \ref{sen_exp} to quantify the degradation of sensitivity in the multimode case under number and homodyne detections.

In the following, to quantify the impact of this residual multimode structure on the phase sensitivity of the SU(1,1) interferometer, the pump FWHM is varied from $1$ to $10\,\mathrm{nm}$, while the waveguide length is fixed at $5\,\mathrm{mm}$. This keeps the nonlinear interaction strength, and hence the total parametric gain, unchanged throughout the analysis. The variation in the spectral purity of the generated photon pairs due to pump bandwidth variation for the designed waveguide is shown in Figure\,\ref{fig_purity}. For each pump bandwidth, the Schmidt coefficients are extracted from the simulated JSA and subsequently used in the sensitivity expressions derived in Section\,\ref{sen_exp}. Since the nonlinear interaction strength is held fixed, changing the pump bandwidth only redistributes the fixed squeezing resource among the Schmidt modes, leading to a degradation in the sensitivity of the SU(1,1) interferometer. This enables systematic quantification of sensitivity degradation in the multimode case under both number- and homodyne-detection schemes.

\subsection{Sensitivity Degradation Analysis}
\label{Sen_degradation_analysis}

\begin{figure*}[t]
    \centering
    \includegraphics[width=1.0\linewidth]{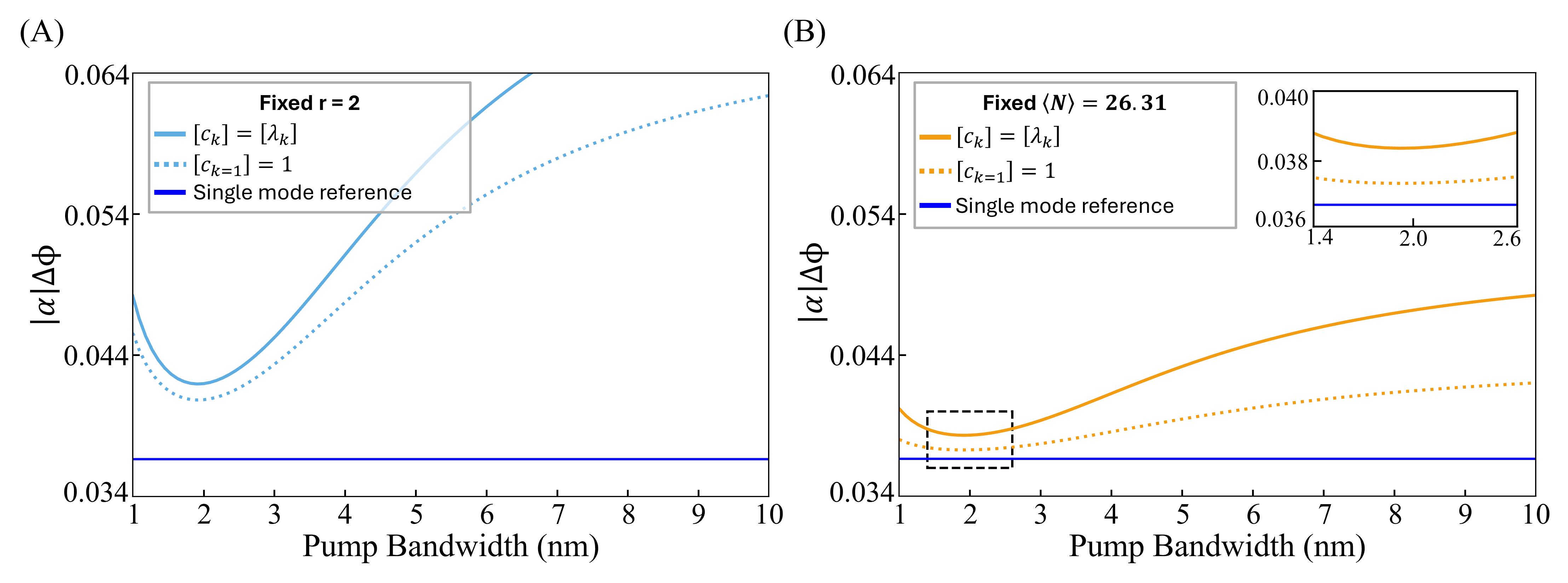}
    \caption{Phase sensitivity obtained using photon-number detection at the idler output port under (A) fixed-$r$ constraint\,(B) fixed- $\langle N \rangle$ constraint, as a function of the pump bandwidth. Sensitivities are shown for both coherent probe distributed according to the Schmidt spectrum ${[c_{k}] = [\lambda_{k}]}$ (full line) and mode-matched to the highest Schmidt-mode ${[c_{k=1}] = 1}$ (dotted line). Sensitivity for the case of a single Schmidt mode is also shown as a reference (thick dark blue full line).} 
    \label{fig5_num_sen}
\end{figure*}

\begin{figure*}[t]
    \centering
    \includegraphics[width=1.0\linewidth]{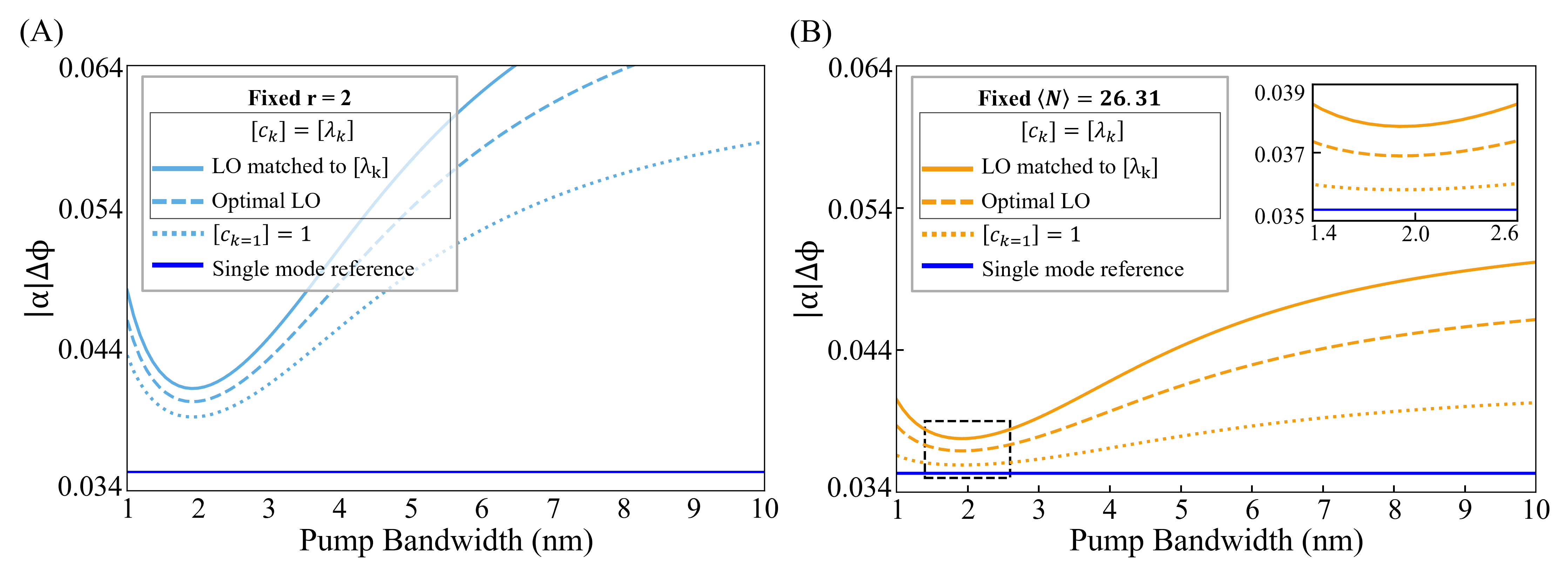}
    \caption{Phase sensitivity using homodyne detection at the signal output port under (A) fixed-$r$\ constraint and (B) fixed- $\langle N \rangle$ constraint. Several LO and input probe configurations are considered. Full line: both the LO and the input probe are matched to the Schmidt spectrum. Dashed line: optimized LO spectral profile. Dotted line: both the LO and the input probe signal spectrum are matched to the strongest Schmidt mode. Thick dark blue full line: single-mode case.}
    \label{fig6_hd_sen}
\end{figure*}

The Schmidt coefficients extracted from the simulated PDC process are now incorporated in the analytical expressions presented in section \ref{sen_exp} (Eqs. \eqref{mm_sen_number} and \eqref{MM_hd}) to quantify the effect of the multimode structure on the phase sensitivity of the SU(1,1) interferometer. To isolate the impact of the Schmidt-mode distribution, the analysis is performed under two complementary resource constraints, each highlighting a distinct physical origin of the degradation of multimode sensitivity. 

The first corresponds to a \textit{fixed total nonlinear gain/total squeezing parameter} ($\mathrm{fixed}\,r$), which, from an experimental perspective, is equivalent to operating at a fixed pump power for a given waveguide length, as discussed earlier. Under this condition, changing the pump bandwidth redistributes the available total squeezing parameter among the Schmidt modes ($r_{k}=r\,\lambda_{k}$) while leaving the overall nonlinear interaction strength unchanged. Any variation in sensitivity originates solely from the redistribution of the squeezing resource. 

The second constraint   (nicknamed fixed $\langle N\rangle$ in the following) adopts a \textit{quantum-resource-based perspective}. It fixes the total number of photons $\langle N \rangle = \sum_{k} 2 \sinh{r\,\lambda_{k}}^{2}$ inside the interferometer that would be spontaneously generated by the first OPA of the interferometer if it were seeded with vacuum for both the signal and idler. Since the spontaneous photon number depends on the Schmidt-mode distribution, varying the pump bandwidth modifies the available quantum resource even when the nonlinear gain is held constant. To eliminate this effect, the squeezing strength $r$ is adjusted for each pump bandwidth to keep the total number of spontaneous photons unchanged. This provides a complementary comparison in which the quantum resource is identical across all Schmidt spectra, allowing the impact of the Schmidt-mode distribution to be examined independently of changes in the total number of generated photons. 

The sensitivity results presented below are obtained using a single-mode squeezing parameter $r=2$, corresponding to the Bogoliubov coefficient $G=3.76$. For the multimode interferometer, the squeezing strength is distributed among the Schmidt modes according to $r_{k}=r\,\lambda_{k}$. Under the fixed spontaneous-photon-number constraint, this corresponds to a total spontaneous photon number of $\langle N \rangle = 26.31$. Choosing a higher squeezing strength ($r > 2$), as reported in other studies \cite{sq_Spasibko2012,sq_Sharapova2020,sq_Hirota2026,sq_Karnik2026}, would have a more detrimental effect on sensitivity degradation for the same Schmidt distribution as shown in the Appendix.

\subsubsection{Number Detection}

The phase sensitivity obtained using photon-number detection at the idler output port (see Figure \ref{fig1_SU11_block_dia}-C) for $\ket{\alpha_{f}}_{s}\,\otimes\,\ket{0}_{i}$ input is presented in Figure \ref{fig5_num_sen} as a function of the pump bandwidth. As discussed in Section \ref{Sen_degradation_analysis}, two resource constraints are considered. Under the fixed-$r$ constraint (see Figure \ref{fig5_num_sen}-A), the pump power is kept constant so that the total nonlinear gain remains unchanged, while varying the pump bandwidth redistributes the squeezing strength among the Schmidt modes according to  $r_{k} = r\lambda_{k}$. Under the fixed- $\langle N \rangle$ constraint (see Figure \ref{fig5_num_sen}-B), the pump power is adjusted for each pump bandwidth to maintain a constant number of spontaneously generated photons inside the interferometer, thereby isolating the influence of the Schmidt-mode distribution from changes in available quantum resource. Two coherent input configurations are also investigated: the dotted lines correspond to a coherent probe mode being matched to the dominant Schmidt mode $({[c_{k=1}] = 1}$), while the full lines correspond to the case of a coherent probe spectrum distributed on the different modes according to the Schmidt distribution $({[c_{k}] = [\lambda_{k}]}$) of the PDC process as shown in Figure \ref{fig4_Schmidt_mode}-D. 

In both cases, the minimum phase sensitivity is obtained for the pump bandwidth that yields the highest spectral purity (Figure\,\ref{fig_purity}): an increase in multimode character progressively degrades the sensitivity. Figure \ref{fig5_num_sen} shows that the degradation in sensitivity is significantly reduced under a fixed $\langle N \rangle$ constraint compared to the case of the fixed-$r$ constraint, indicating that a substantial fraction of the penalty observed in the  fixed-$r$ case originates from the decrease in generated photon resource. Furthermore, in both cases, the sensitivity degradation can be partially mitigated by matching the coherent signal probe spectrum to the dominant Schmidt mode (Figure\,\ref{fig5_Schmidt_mode_profiles}). In particular, the inset in Figure \ref{fig5_num_sen}-B shows that the best sensitivity achieved in this case becomes very close to the one that would be achieved in the single-mode situation. 

The observations made here based on the simulations of our PP-TFLN parametric amplifiers are in complete agreement with the discussion in Section \ref{num_theory}, confirming that the multimode penalty in the case of photon-number detection arises solely from the redistribution of the nonlinear resource among Schmidt modes.

\subsubsection{Homodyne Detection}

The phase sensitivities obtained using homodyne detection are presented in Figure \ref{fig6_hd_sen} under the fixed-$r$ (Figure \ref{fig6_hd_sen}-A) and fixed-$\langle N \rangle$ (Figure \ref{fig6_hd_sen}-B) resource constraints. Two LO configurations are considered: an LO whose spectral profile follows the Schmidt distribution of the gain and an optimal LO given by Eq. \eqref{opt_lo}. As with the photon-number measurement, the minimum phase sensitivity is achieved for the pump bandwidth that yields the highest spectral purity (Figure\,\ref{fig_purity}), regardless of the LO spectral configuration and type of constraint. 

In contrast to number detection, the performance of homodyne detection is also governed by the spectral overlap between the signal field and the local oscillator, with both expressed in the Schmidt basis. From Figure \ref{fig6_hd_sen}, we observe that the optimally shaped LO consistently outperforms the LO shaped according to the Schmidt distribution under both resource constraints. Finally, a minimum sensitivity degradation is observed when the coherent probe and the LO are both mode-matched to the dominant Schmidt mode, as shown by the inset in Figure \ref{fig6_hd_sen}-B. These observations are in excellent agreement with the discussion in Section \ref{hd_theory}, confirming that the optimal LO shaping suppresses the coherence penalty, while the remaining sensitivity degradation is determined by the redistribution of the available nonlinear resource among the Schmidt modes. Finally, the main conclusion here is that the sensitivity degradation induced by the multimode nature of the PDC process can be, to a large extent, mitigated by i) using homodyne detection; ii) matching the input signal spectrum to the strongest Schmidt mode, and iii) optimizing the LO spectrum along  Eq. \eqref{opt_lo}.

\section{Conclusion}
In this work, we have developed a comprehensive theoretical framework to evaluate the phase sensitivity of a realistic multimode SU(1,1) interferometer, based on the Schmidt-mode representation of the PDC source. Focusing on the widely studied coherent-vacuum input state, we derived analytical expressions for the interferometric phase sensitivities for number and homodyne detections, the two most experimentally relevant measurement schemes for SU(1,1) interferometry. We showed that, under the assumptions of identical OPAs and a mode-independent phase shift, the multimode interferometer can be rigorously described as a tensor product of the independent Schmidt-mode interferometer, providing a convenient framework for analyzing realistic waveguide-based OPAs. Furthermore, we demonstrated that the physical origin of the multimode penalty depends on the detection scheme employed. Number detection is limited by the redistribution of the available nonlinear resource among the Schmidt modes. In addition to the nonlinear resource distribution, homodyne detection suffers from a coherence penalty arising from the coherent superposition of the measured Schmidt-mode quadratures defined by the LO. This coherence penalty can thus be significantly reduced through appropriate local-oscillator shaping.

To demonstrate the practical relevance of the proposed framework, we designed and numerically investigated a PP-TFLN waveguide-based OPA engineered to generate a nearly factorizable biphoton state with controllable multimode characteristics described by its Schmidt decomposition. By incorporating the extracted Schmidt distribution directly into the analytical sensitivity expressions, we established a direct connection between the realistic waveguide design and interferometric performance. The numerical results under both fixed nonlinear-gain and fixed spontaneous-photon-number resource constraints quantitatively reveal how the OPA's spectrally multimode nature determines the achievable phase sensitivity.

Overall, this work bridges the gap between the mature field of SPDC spectral engineering and the phase-sensitivity analysis of SU(1,1) interferometers. The theoretical framework developed here is readily applicable to any waveguide platform whose SPDC process can be characterized through its Schmidt decomposition. The PP-TFLN waveguide design presented in this work provides a practical route toward implementing a high-performance multimode SU(1,1) interferometer for quantum-enhanced optical phase sensing.

\bmsubsection*{Acknowledgments}
Sonu Jana acknowledges the financial support received from the Ministry of Education (MoE), Government of India, through the Prime Minister's Research Fellowship (PMRF) program. Syamsundar De acknowledges the financial support received from the Ministry of Electronics and Information Technology (MeitY), Government of India, under the project "National Center for Quantum Accelerated Chips using Lithium Niobate (NCQAC)." This work was supported by the Physics Graduate School of Universit\'e Paris-Saclay (project NORI-MAQI) and the Institut des Sciences de la Lumière of Universit\'e Paris-Saclay (project ROLLMOPS). Fabien Bretenaker acknowledges support from IIT Kharagpur in the form of repeated invitations and an adjunct professor position.

\bmsubsection*{Conflicts of Interest}

The authors declare no conflicts of interest.

\bmsubsection*{Data Availability Statement}

No such dataset was generated during this study.


\newpage
\bmsubsection*{Appendix}
\begin{figure}[h]
    \centering
    \includegraphics[width=1\linewidth]{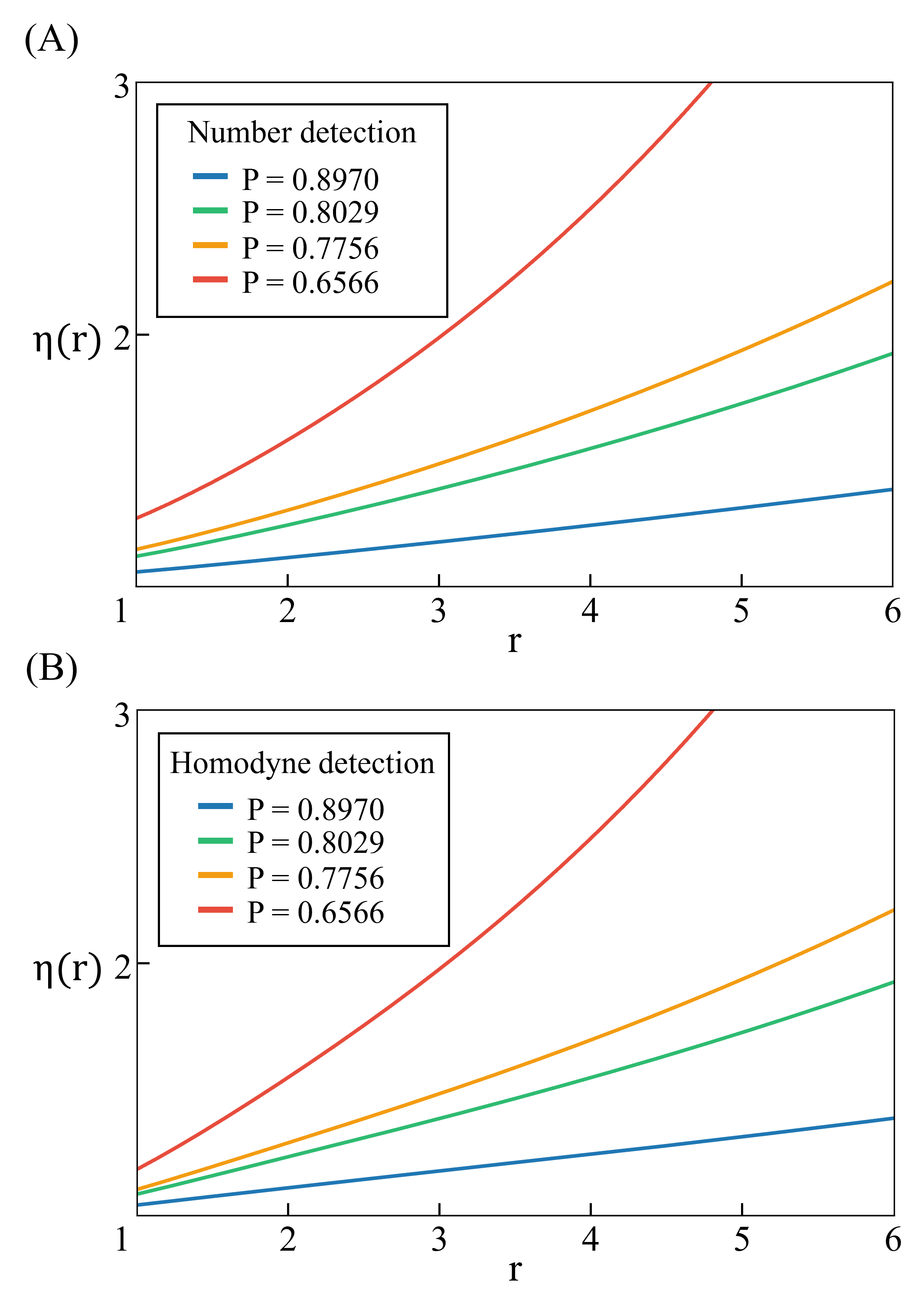}
    \caption{Ratio of the multimode to single -mode phase sensitivity $\eta(r)$, as a function of the total squeezing strength $r$. (A) Number detection. (B) Homodyne detection. The ratio is plotted as a function of purity (P). The increasing deviation from unity as $r$ increases indicates that the degradation of multimode sensitivity becomes progressively more significant at higher nonlinear gain.} 
    \label{fig:sen_ratio}
\end{figure}
In this Appendix, we examine how the multimode sensitivity degradation depends on the total squeezing strength ($r$). To quantify this effect, we defined the sensitivity degradation penalty as a ratio of the phase sensitivity obtained with multimode and corresponding single-mode SU(1,1) interferometers,
\begin{equation}
    \eta(r) = \frac{\Delta \varphi_{\mathrm{MM}}}{\Delta \varphi_{\mathrm{SM}}}
\end{equation}

where $\eta (r) = 1$ corresponds to the single-mode limit. Here, $\eta (r) > 1$ indicates a degradation in sensitivity arising from the multimode nature of the interferometer. Figure \ref{fig:sen_ratio} shows the variation of the sensitivity degradation penalty as a function of $r$ in the case of number and homodyne detection for different values of purity ($P$).

The observed sensitivity-degradation penalty arises from the distribution of the total squeezing strength across the different Schmidt modes. In the multimode interferometer, the squeezing parameter for the $k^{\mathrm{th}}$ Schmidt mode is $r_{k}=r\lambda_{k}$. The amplitude gain, $G_{k} = \cosh{r_{k}}, g_{k} = \sinh{r_{k}}$ of the interferometer depends nonlinearly on $r$. As the total nonlinear gain increases, the gain of each Schmidt mode increases more slowly than the corresponding single-mode case, resulting in a more pronounced redistribution of the nonlinear gain among the different Schmidt modes.


\end{document}